\documentclass{aa}
\usepackage{graphicx}
\usepackage{txfonts}
\usepackage{natbib}
\begin{document}
   \title{A new determination of the \textit{INTEGRAL}/IBIS \\point source location accuracy}

   \subtitle{}

   \author{S. Scaringi\inst{1,2}\fnmsep\thanks{email: S.Scaringi@astro.ru.nl}
          \and
          A.J. Bird\inst{1}
          \and
          A.B. Hill\inst{3}
          \and
          D.J. Clark\inst{1}
          \and
          V.A. McBride\inst{1}
          \and
	  A.J. Dean\inst{1}
          \and
          A. Bazzano\inst{4}
          \and
          L. Natalucci\inst{4}
          \and
          J.B. Stephen\inst{5}
          }

   \institute{Department of Astrophysics, IMAPP, Radboud University Nijmegen, P.O. Box 9010, 6500 GL Nijmegen, The Netherlands \\
         \and
             School of Physics and Astronomy, University of Southampton, Southampton, SO17 1BJ, U.K.\\
         \and
             Laboratoire d'Astrophysique de Grenoble, UMR 5571 CNRS, BP 53, 38041 Grenoble, France\\
         \and 
             INAF-IASF Roma, Via Fosso del Cavaliere 100, I-00133, Roma, Italy\\
         \and
	     INAF-IASF Bologna, Via P. Gobetti 101, I-40129, Bologna, Italy
             }

   \date{Received ...; accepted ...}

 
  \abstract
   {}
   {To determine the Point Source Location Accuracy (PSLA) for the INTEGRAL/IBIS telescope based on analysis of archival in-flight data.}
   {Over 40000 individual pointings (science windows) of INTEGRAL/IBIS data were analysed using the latest Off-line Science Analysis software, version 7.0. Reconstructed source positions were then compared against the most accurate positions available, determined from focusing X-ray telescopes. Since the PSLA is a strong function of source detection significance, the offsets from true position were histogrammed against significance, so that the 90\% confidence limits could be determined. This has been done for both sources in the fully coded field of view (FCFOV) and partially coded field of view (PCFOV).}
   {The PSLA is found to have improved significantly since values derived from early mission data and software for both FCFOV and PCFOV.}
   {This result has implications for observers executing follow-up programs on IBIS sources since the sky area to be searched is reduced by $>$50\% in some cases.}


   \maketitle
%
%

\section{Introduction}
The \textit{INTEGRAL} satellite (\citealt{winkler}), launched in October 2002, is an ESA space mission specifically designed to study the gamma-ray sky. In particular the IBIS (Imager on Board of the \textit{INTEGRAL} Satellite) telescope (\citealt{ubertini03}) is the main hard X-ray/soft gamma-ray coded aperture imaging instrument (\citealt{goldwurm03}), and is responsible for surveying and cataloguing the sky above 17 keV. Here we discuss the point source location accuracy (PSLA) of the ISGRI low energy detector (15-1000 keV) of IBIS (\citealt{lebrun03}). 

Due to the continuing necessity to follow-up the growing unidentified \textit{INTEGRAL} source population in other wavebands (particularly optical and infrared), it is of great interest to assure the correctness of the IBIS/ISGRI PSLA. In particular, it is hoped that with the release of new and updated Off-line Science Analysis (OSA 7.0) software  the PSLA could have improved substantially compared to the already published estimates based on early mission data and software releases \citep[][constructed the PSLA based upon OSA 3.0]{gros}. Any improvement in the PSLA is important in reducing the chance of random or multiple source associations in other wavebands.


\section{Refining the PSLA error radius}

In order to empirically determine the PSLA of the IBIS/ISGRI telescope, we can extract the positions of objects from the IBIS/ISGRI Science Windows\footnote{An IBIS/ISGRI Science Window refers to one single \textit{INTEGRAL} pointing of about 2000 seconds} (ScWs) and compare these with their best known positions. This can then be used to estimate the $90\%$  error offset as a function of detected significance, allowing us to define the $90\%$  PSLA. In particular, the IBIS/ISGRI telescope (and coded mask telescopes in general) PSLA depends strongly on detection significance, but also on the position of the source within the field of view. We therefore require for our analysis a set of sources spanning a wide range of significances and off-axis positions in order to allow our analysis to be useful for all detected IBIS/ISGRI sources. 

We begin by compiling a list of sources with good positions. It is best at this stage to be conservative and only select sources where accurate nominal positions are known rather than to bias our sample by also including sources with large nominal error radii, like many newly discovered \textit{INTEGRAL} sources. To do this we take the latest \textit{INTEGRAL} General Reference Catalog (Version 30, \citealt{ebisawa}, \texttt{http://isdc.unige.ch/Data/cat/latest/}) and make a selection on the error radius, selecting those sources with an error less than 30$''$; at this level, the error on the true position should give a negligible contribution to the measured offsets to the IBIS position.  Thus the total number of objects used in our sample is 332, spanning a wide range of detection significances and off-axis angles. This number might seem small when compared to the 721 sources detected in \cite{cat4}, however we note that many objects in that catalog are newly discovered \textit{INTEGRAL} sources for which X-ray follow-up is not yet available, and that therefore have relatively large nominal error radii.

\begin{figure}[tbhp]
   \centering
 	\includegraphics[width=.45\textwidth, height=0.22\textheight]{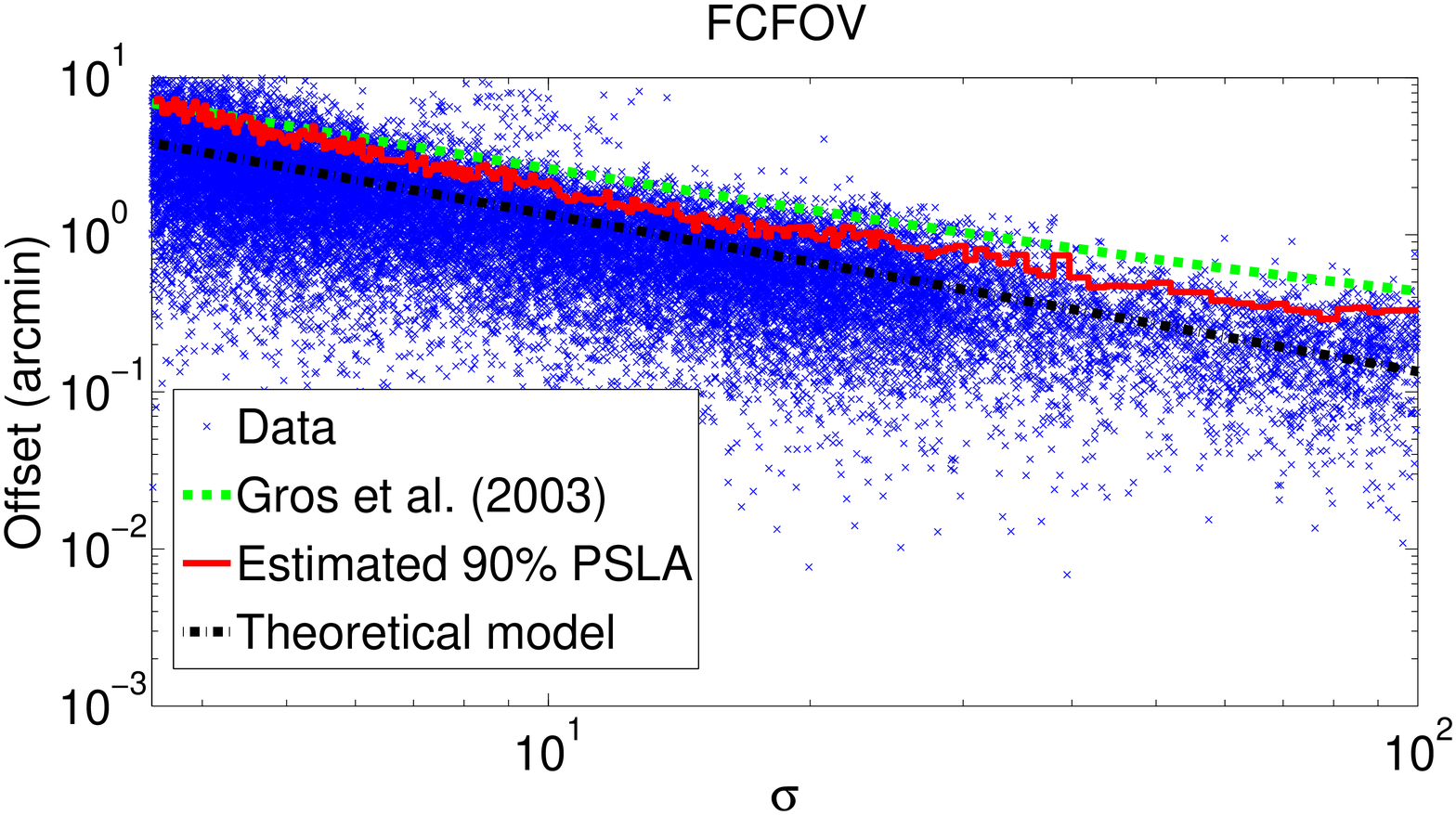}
  	\includegraphics[width=.45\textwidth, height=0.22\textheight]{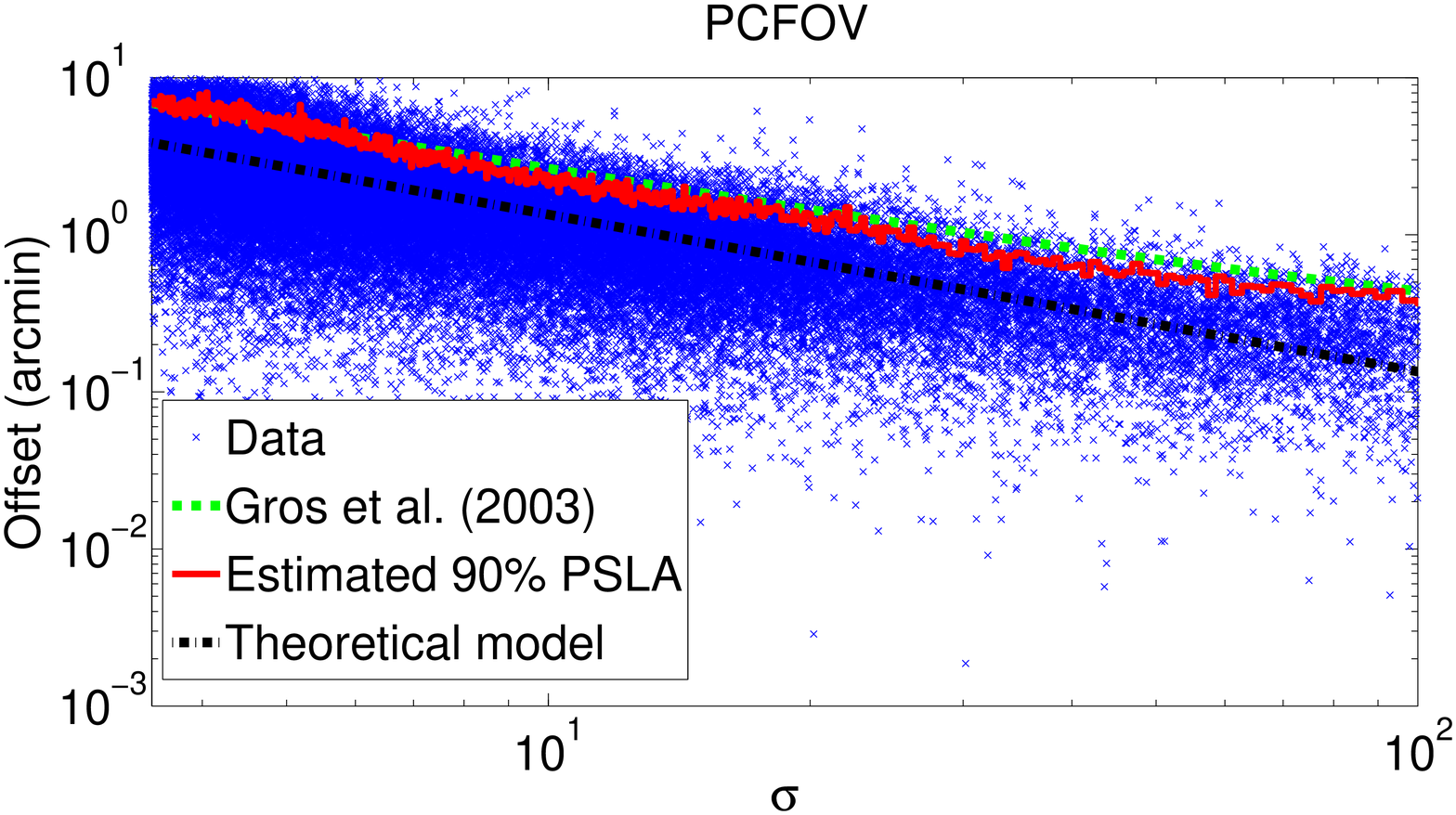}
   \caption{Measured offset from true position (datapoints) as a function of detection significance for IBIS/ISGRI detected sources in the (upper) fully coded field of view and (lower) partially coded field of view. Also plotted is the estimated $90\%$ PSLA from \cite{gros} (dashed line), our estimated $90\%$ PSLA (solid line) and the telescope theoretical PSLA from \cite{goldwurm01} (dashed-dotted line).}
   \label{fig:PSLA}
\end{figure}

After performing an imaging analysis with the OSA 7.0 pipeline, we inspected all available ScW images and extracted the fitted positions which resulted from the image deconvolution; these are the columns RA\_FIN and DEC\_FIN from the \mbox{isgri\_sky\_res} file for all pointings where any of the 332 objects was present. The dataset was divided into fully coded field of view (FCFOV) and partially coded field of view (PCFOV).  The IBIS coded aperture mask has a field of view of 30\degr, the FCFOV is the central 9$\degr\times$9$\degr$ region; everything outside of this region constitutes the PCFOV. We then determine the offset as a function of detection significance in both FCFOV and PCFOV, for all objects in question, allowing us to estimate the $90\%$ PSL confidence by adaptively binning the measured offsets in significance in order to have the same statistics for all bins (100 measurements per bin). Around 25000 and 75000 individual offset measurements are used in the FCFOV and PCFOV analyses respectively. The results are shown in Figure \ref{fig:PSLA} with the solid line representing the 90\% confidence limit, whilst the dashed line shows the result from \cite{gros} and the dashed-dotted line shows the theoretical PSLA as defined by \cite{goldwurm01}. Note that the theoretical PSLA of \cite{goldwurm01} applies only for on-axis sources, whilst the estimated $90\%$ PSLA from \cite{gros} is derived for sources within 14 degrees of the telescope axis.

\begin{figure}[tbhp]
   \centering
   \includegraphics[width=0.9\columnwidth]{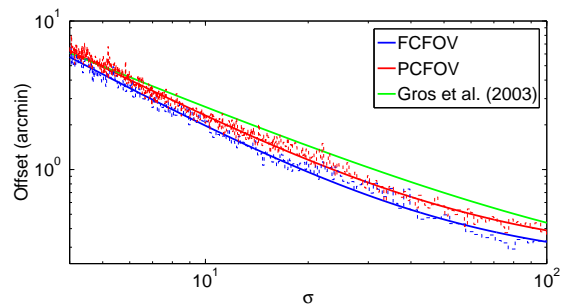}
   \caption{Fits to the OSA 7.0 $90\%$ PSLA in FCFOV and PCFOV. In each case, the solid line represents the best fit. Also plotted is the $90\%$ PSLA estimate of \cite{gros}.}
   \label{fig:PnF}
\end{figure}

In order to fit the estimated $90\%$ PSLA we have used the form $y=ax^{c}+b$ (the same as used by \citealt{gros}), and applied the same weights for all bins as they contain the same number of observations. Similarly for completeness we also estimated the $65\%$, $95\%$ and $99\%$ PSLA. The fit parameters for all estimated PSLA are shown in Table \ref{all_fits}, whilst we display only the $90\%$ PSLA in Figure \ref{fig:PnF}.


\begin{table}
\begin{centering}
\caption{Estimated PSLA fit parameters for the FCFOV and the PCFOV.}
\begin{tabular}{c c c}
\hline
  & FCFOV & PCFOV \\[0.5ex]
\hline
$65\%$ & $a=17.65$ & $a=18.78$ \\
       & $b=0.14$ & $b=0.17$ \\
       & $c=-1.19$ & $c=-1.16$ \\
\hline
$90\%$ & $a=31.1$ & $a=31.1$ \\
       & $b=0.23$ & $b=0.25$ \\
       & $c=-1.25$ & $c=-1.18$ \\
\hline
$95\%$ & $a=36.7$ & $a=34.8$ \\
       & $b=0.25$ & $b=0.26$  \\
       & $c=-1.24$ & $c=-1.14$  \\
\hline
$99\%$ & $a=36.4$ & $a=34.6$ \\
       & $b=0.14$ & $b=0.19$ \\
       & $c=-1.04$ & $c=-0.97$ \\
\hline
\end{tabular}
\\
\label{all_fits}
\end{centering}
\end{table}


\section{Comparison to previous results}
\begin{figure}[htbp]
   \centering
   \includegraphics[width=.45\textwidth, height=0.22\textheight]{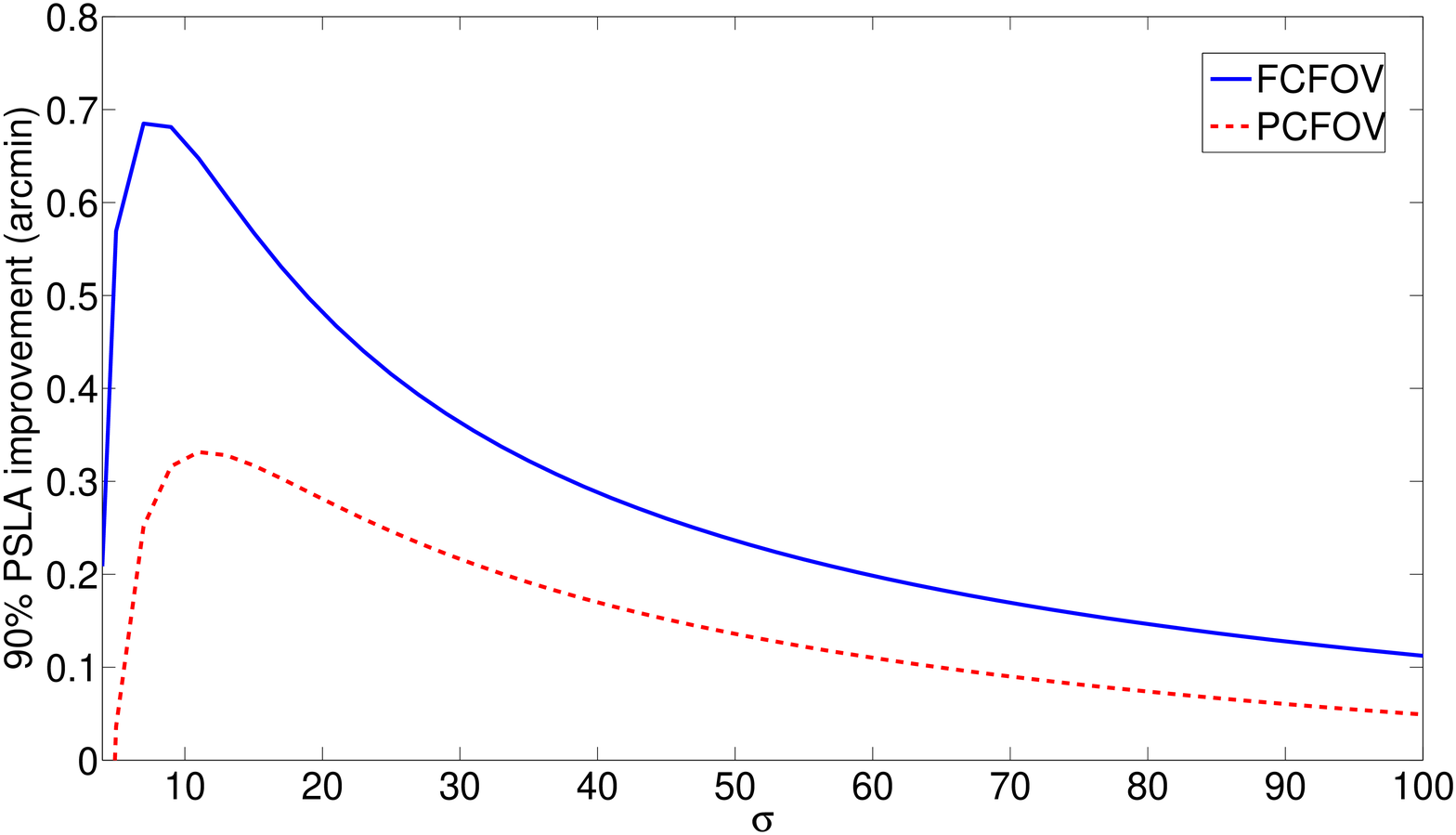}
   \caption{$90\%$ PSL error radius improvement between the old \cite{gros} and the new (this work) estimates in arcminutes. }
   \label{fig:IMP1}
\end{figure}

\begin{figure}[htbp]
   \centering
   \includegraphics[width=.45\textwidth, height=0.22\textheight]{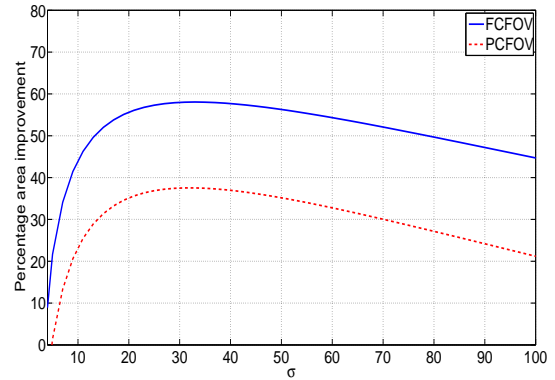}
   \caption{Percentage sky search area improvement between the old \cite{gros} and the new (this work) $90\%$ error radii.}
   \label{fig:IMP2}
\end{figure}

These empirical fits can now be used in order to estimate the improvement in radius and area between the previously published fit of \cite{gros} and our own estimates. To this end, Figure \ref{fig:PnF} shows once again our fits for the 90\% PSLA for the FCFOV and PCFOV and the \cite{gros} result. To demonstrate the decrease in error radius between the \cite{gros} result and our new updated PSLA we simply subtract the two fits in order to display the error radius improvement. This is shown in Figure \ref{fig:IMP1}, where it can be seen that the biggest reduction in radius occurs at approximately $10\sigma - 15\sigma$. The potential for reducing false matches when performing follow-up observations of such objects in other wavebands is however best shown in Figure \ref{fig:IMP2}, where the percentage area improvement of the 90\% error circle is shown. This is defined to be the area subtended by our $90\%$ PSLA as a function of significance divided by the area subtended by the $90\%$ PSLA estimate of \cite{gros}. Here the biggest reduction occurs in the range $20\sigma - 25\sigma$, where the area to inspect reduces by $\approx 50\%$ from the previous \cite{gros} result. This would be a typical significance for a transient detection in a single ScW of a source of $\sim 250$mCrab. We therefore predict that these transient detections will benefit the most from this improved 90\% PSLA, where the probability of false matches for follow-up observations will be drastically reduced.

\textit{INTEGRAL}/IBIS has also detected and classified a large number of persistent sources such as AGN, which obtain their peak significance after mosaicking many individual ScWs. In this case it is not always easy to establish how many pointings were used where the source candidate was in the FCFOV or the PCFOV, and so it is not trivial to establish which 90\% PSLA to use. In these cases we suggest a conservative approach and use the PCFOV PSLA estimate.

We have also carried out the same analysis for data reduced with and older version of the analysis software, OSA 5.0, and find after fitting the $90\%$ PSLA, that an improvement on the estimated error circle size had already been achieved with OSA 5.0 compared to the \cite{gros} result based on OSA 3.0; however with the latest version of the software release, OSA 7.0, the improvement is even more evident. This result also indicates that a re-analysis of archival data with the latest software will often yield far better source positions, and may allow observers to distinguish between multiple candidate counterparts far better than previously.

\section{Conclusions}

This work has shown that the IBIS/ISGRI PSLA (ie the size of the 90\% error circle) has dramatically improved since the last study of \cite{gros} and this can be attributed to improvements in the {\em OSA} software. The main implications of this result will be for follow-up observations, particularly in the optical and infrared wavebands, of the \textit{INTEGRAL} unidentified source population ($\approx30\%$  of objects present in the latest IBIS/ISGRI catalogue release by \citealt{cat4}). This improved error radius will greatly enhance the identification of \textit{INTEGRAL} counterparts.

\begin{acknowledgements}
Based on observations with INTEGRAL, an ESA project funded by member states (especially the PI countries: Denmark, France, Germany, Italy, Switzerland, Spain), Czech Republic and Poland, and with the participation of Russia and the USA. 

AJB and DJC acknowledge support from STFC award ST/G004196/1. ABH acknowledges support from the European Community via contract ERC-StG-200911. AB, LN and JS acknowledge ASI/INAF grant I/008/07/0.

This research has made use of NASAs Astrophysics Data System Bibliographic Services, of the SIMBAD data base, operated at the CDS, Strasbourg, France, as well as of the NASA/IPAC Extragalactic Data base, which is operated by the Jet Propulsion Laboratory, California Institute of Technology, under contract with the National Aeronautics and Space Administration.
\end{acknowledgements}

\bibliographystyle{aa}
\bibliography{PSL_paper}

\end{document}